\documentclass[a4paper,11pt]{article}
\pdfoutput=1 

\usepackage{jcappub} 

\usepackage{amssymb,amsmath}
\usepackage{verbatim,xspace}

\newcommand{\ro}[1]{\ensuremath{\textrm{#1}}}
\newcommand{\ten}[1]{\ensuremath{\times 10^{#1}}}

\newcommand{\Msol}{\ensuremath{M_{\odot}}\xspace}
\newcommand{\Lsol}{\ensuremath{L_{\odot}}\xspace}

\newcommand{\df}{\ensuremath{~ \ro{d}}}

\newcommand{\mPr}{\ensuremath{m_\ro{p}}\xspace}

\newcommand{\mR}{\ensuremath{m_\ro{R}}\xspace}
\newcommand{\mion}{\ensuremath{m_\ro{i}}\xspace}
\newcommand{\cC}{\ensuremath{\mathcal{C}}\xspace}
\newcommand{\Deu}{\ensuremath{[\ro{D}/\ro{H}]}\xspace}
\newcommand{\alG}{\ensuremath{\alpha_{\!_G}}\xspace}

\usepackage[T1]{fontenc} 

\title{Binding the Diproton in Stars: Anthropic Limits on the Strength of Gravity}

\author[a]{Luke A. Barnes}

\affiliation[a]{Sydney Institute for Astronomy \\
School of Physics, A28 \\
The University of Sydney \\
NSW 2006, Australia}

\emailAdd{L.Barnes@physics.usyd.edu.au}

\abstract{We calculate the properties and investigate the stability of stars that burn via strong (and electromagnetic) interactions, and compare their properties with those that, as in our Universe, include a rate-limiting weak interaction. It has been suggested that, if the diproton were bound, stars would burn $\sim10^{18}$ times brighter and faster via strong interactions, resulting in a universe that would fail to support life. By considering the representative case of a star in our Universe with initially equal numbers of protons and deuterons, we find that stable, ``strong-burning'' stars adjust their central densities and temperatures to have familiar surface temperatures, luminosities and lifetimes. There is no ``diproton disaster''. In addition, strong-burning stars are stable in a much larger region of the parameter space of fundamental constants, specifically the strength of electromagnetism and gravity. The strongest anthropic bound on stars in such universes is not their stability, as is the case for stars limited by the weak interaction, but rather their lifetime. Regardless of the strength of electromagnetism, all stars burn out in mere millions of years unless the gravitational coupling constant is extremely small, $\alG \lesssim 10^{-30}$.}

\begin{document}
\maketitle
\flushbottom

\section{Introduction}
\label{sec:intro}

Beginning in the 1970's, physicists have noted the extreme sensitivity of the life-permitting qualities of our universe to the values of many of its fundamental constants and cosmological parameters. Seemingly small changes to the free parameters of the laws of nature as we know them would have dramatic, uncompensated and detrimental effects on the ability of a universe to support the complexity needed by physical life forms. We have elsewhere reviewed the scientific literature on the fine-tuning of the universe for intelligent life \cite{Barnes2012}; here are some illustrative examples.
\begin{itemize}
\item The existence of \emph{any} structure in the universe places stringent bounds on the cosmological constant. Compared to the ``natural'' scale of the vacuum energy of quantum fields (approximately $\pm$ the Planck scale), the range of values that permit gravitationally bound structures is no more than one part in $10^{110}$ \cite{1987PhRvL..59.2607W,1995MNRAS.274L..73E,2007MNRAS.379.1067P,2006PhRvD..73b3505T}.
\item A universe with structure also requires a fine-tuned value for the primordial density contrast $Q$. Too low, and no structure forms. Too high and galaxies are too dense to allow for long-lived planetary systems, as the time between disruption by a neighbouring star is too short. This places the constraint $10^{-6} \lesssim Q \lesssim 10^{-4}$ \cite{Tegmark1998}. These arguments have been recently refined in \cite{Adams2015}.
\item The existence of any atomic species and chemical processes whatsoever places tight constraints on the  masses of the fundamental particles and the strengths of the fundamental forces. For example, Figure 4 of \cite{2007PhRvD..76d5002B} shows the effect of varying the masses of the up and down quark, finding that chemistry-permitting universes are huddled in a small shard of parameter space which has area $\Delta m_\textrm{up} ~ \Delta m_\textrm{down} ~/~ m^2_\textrm{Planck} \approx 10^{-40}$, where $m_\ro{up}$ ($m_\ro{down}$) is the mass of the up (down) quark, and $m_\ro{Planck} = \sqrt{\hbar c/G}$ is the Planck mass, which combines the (reduced) Planck's constant ($\hbar$), the speed of light ($c$) and Newton's gravitational constant ($G$). Similarly, the stability of free protons requires $\alpha \lesssim (m_\textrm{down} - m_\textrm{up})/141$ MeV $\approx 1/50$ \cite{2000RvMP...72.1149H,2008PhRvD..78c5001H}, where $\alpha = e^2 / \hbar c$ is the fine structure constant.
\end{itemize}
Note that these constraints are all multi-dimensional; we have quoted one-dimensional bounds for simplicity. Constraints in multiple dimensions of parameter space are presented in \cite{Barnes2012}.

Dyson \cite{Dyson1971} was the first to note what seems to be an outstanding case of fine-tuning regarding the diproton ($^2$He):
\begin{quote}
The crucial difference between the sun and a bomb is that the sun contains ordinary hydrogen with only a trace of the heavy hydrogen isotopes deuterium and tritium, whereas the bomb is made mainly of heavy hydrogen. Heavy hydrogen can burn explosively by strong nuclear interactions, but ordinary hydrogen can react with itself only by the weak interaction process. In this process two hydrogen nuclei (protons) fuse to form a deuteron (a proton and a neutron) plus a positron and a neutrino. The proton-proton reaction proceeds about $10^{18}$ times more slowly than a strong nuclear reaction at the same density and temperature. It is this weak-interaction hangup that makes ordinary hydrogen useless to us as a terrestrial source of energy. The hangup is essential to our existence, however \ldots [W]ithout this hangup we would not have a sufficiently long-lived and stable sun.
\end{quote}

This is all the more remarkable given the seemingly small change to the constants of nature that would be required to bind the diproton and remove the weak-interaction hangup in stars that burn ordinary hydrogen. The constants of our universe are in a narrow range in which the deuteron is stable but the diproton and dineutron are unstable. In a nuclear potential well that is $\sim 40$ MeV deep, the diproton is unbound by a mere 0.092 MeV, while the deuteron is bound by 2.2 MeV. Increasing the strength of the strong force by 6\%  would bind the diproton, while decreasing its strength by 4\% would unbind the deuteron, which is the first product of stellar nuclear burning \cite{1991A&A...243....1P,Davies1972,Davies1983,1986acp..book.....B,1998AnPhy.270....1T,2008PhTea..46..285C}. Barr and Khan \cite{2007PhRvD..76d5002B} calculate the equivalent limits on the masses of the light quarks, finding that the diproton is bound if $m_\ro{up} + m_\ro{down} < 0.75~ (m_\ro{up,0} + m_\ro{down,0}) = 5.3$ MeV, where a subscript zero refers to the value of the quark mass in our Universe: $m_\ro{up,0} = 2.3$ MeV and $m_\ro{down,0} = 4.8$ MeV \cite{Olive2014}. Further, the deuteron is strongly unbound ($D \rightarrow p + n$) if $m_\ro{up} + m_\ro{down} > 1.4~ (m_\ro{up,0} + m_\ro{down,0}) = 9.94$ MeV. If all $A = 2$ nuclei (diproton, dineutron and deuteron) were strongly unstable, then nuclear reactions inside stars would require weak three-body reactions, and thus very high central temperatures and densities. The stellar window would narrow significantly if not close altogether, as small proto-stars would fail to ignite \cite{2006PhRvD..74l3514H}. 

It has also been argued \cite{Dyson1971,1986acp..book.....B} that a bound diproton would result in all the hydrogen in the universe being consumed in the first few minutes after the big bang. All the matter in the universe would have been burned to helium, leaving no hydrogen for water, organic compounds and ``normally long-lived stars'' \cite{Dyson1971}. ``If the di-proton existed, \emph{we} would not!'', say Barrow and Tipler \cite{1986acp..book.....B} .

However, more detailed calculations of big bang nucleosynthesis (BBN) \cite{2009PhRvD..80d3507M,2009JApA...30..119B} have shown that the binding of the diproton is not sufficient to ensure that hydrogen is completely burnt in the early universe. The reason is that the diproton and dineutron are always less tightly bound than the deuteron, and so diproton production must wait until the universe is sufficiently cool; this is the equivalent of the ``deuterium bottleneck'' \cite{Mukhanov2005}. Even in universes in which the strong force binds the diproton, by the time it is stable against photodisintegration by background radiation, the rate of diproton production can be very low due to the freezing out of the proton-proton reaction. Further reactions can also be inefficient, so that $^4$He production is unaffected; the diprotons decay by the weak-interaction into deuterons. The net result is a larger deuteron abundance than standard BBN. Only when the strength of the strong force in increased by $\sim$50\% does BBN produce a mostly $^4$He universe.

Philips \cite{PhillipsA.C.1999} and Bradford \cite{2009JApA...30..119B} have argued that long-lived stars may be possible in universes in which the diproton is bound. Because stars are thermodynamic systems, the nuclear burning rate is determined by the overall stability of the star. Energy is supplied at the centre of the star to maintain thermal pressure to balance the crush of gravity, countering the leaking of energy as photons escape from the star's outer surface. The density and temperature of the centre of the star will adjust to maintain the  energy production rate required for stable burning. In particular, simple, order-of-magnitude stellar models \cite{1979Natur.278..605C,1986acp..book.....B} provide the following estimate for the main-sequence lifetime of stars,
\begin{equation}
t_* = \epsilon_\ro{nuc} \left( \frac{\alpha \mPr} {m_\ro{e}} \right)^2 \left( \frac{M_*} {M_0} \right)^{-2} \alG^{-3/2} ~,
\end{equation}
where $\epsilon_\ro{nuc}$ is the fraction of the rest energy of the star that can be released in nuclear reactions, $m_\ro{e}$ and $\mPr$ are the electron and proton mass respectively, $\alG = (\mPr/m_\ro{Planck})^2$ is the gravitational coupling constant, $M_*$ is the mass of the star, and $M_0 = \alG^{-3/2} \mPr$ is a characteristic stellar mass. Note that this estimate does not depend on the microphysics of the star's central nuclear reaction.

In light of these considerations, we will use a model for stars --- including hydrodynamic equilibrium and heat balance --- to consider the stability and lifetime of stars that burn by strong and electromagnetic interactions, without a weak interaction hangup. In particular, we will consider a star that is powered by equal numbers protons and deuterons, so that we can use the parameters of physics in our Universe. Future investigations may choose to consider the effects of smoothly varying the strength of the strong force; however, one of our conclusions in this work is that stars are remarkably oblivious to such changes in their microphysical processes.

This paper is organized as follows. Section \ref{S:StarModel} outlines the stellar model of Adams (2008) \cite{2008JCAP...08..010A}, and the parameters that we will use to model strong-burning stars. We use the model to calculate the minimum and maximum masses of stars, such that they are massive enough to ignite nuclear reactions, but not so massive that they are subject to the instabilities of photon-pressure supported gravitational systems. We also calculate where stars are stable in parameter space. Section \ref{S:Dburn} applies our model to strong-burning stars, calculating the relations between their mass, core temperature, surface temperature, luminosity, and lifetime. We show the regions in parameter space where strong-burning stars are stable, and also map the lifetimes and surface temperatures of stars in other universes. Our results are discussed in Section \ref{S:Discuss}.

\section{Stellar Model of Adams (2008)} \label{S:StarModel}
\subsection{Model Equations}
Our stellar model follows the formalism of Adams (2008) \cite{2008JCAP...08..010A}. We will summarize the key equations in this section. Readers familiar with Adams's model should note our choice of the free parameters of the model in Table \ref{tab:table_params} and then can skip to Section \ref{S:Dburn}. In short, a system of four coupled differential equations describes the structure of the star: force balance (hydrostatic equilibrium), conservation of mass, heat transport and energy generation. To these we add the equation of state, stellar opacity, and the relationship between the nuclear reaction rate, temperature and density.

\begin{table*}
\centering
\begin{tabular}{|c|c|c|c|}
\hline
Model Parameter & Symbol [units] 								 & H-burning 				& D-burning \\ 
\hline
Polytropic index    & $n$ 											  & 3/2 						& 3/2 \\
Gamow energy      & $E_G$ [keV] 								  & 493.1						& 657.5 \\
Burning efficiency & $\epsilon_{\ro{nuc}}$ 					  & 0.0071 					& 0.0045 \\
Hydrogen mass fraction & $X_H$							      & 0.71           				& 1/3       \\
Helium mass fraction & $Y_\ro{He}$							  & 0.29           				& 0        \\
Deuterium number fraction &  \Deu  						  	  & negligible   				& 1        \\
Nuclear parameter & \cC [cm$^5$ s$^{-3}$ g$^{-1}$] & 2 \ten{4} 				& $2.3 \ten{21}$\\
Central opacity	 	& $\kappa_0$ [cm$^2$ g$^{-1}$]    & 1.4 					        & 0.265 \\
Critical gas/total pressure & $f_g$								  & 1/2						    & 1/2 \\
Average particle mass 		& $\bar{m}$ 						  & 0.61						& 3/4 \mPr \\
Mass per free electron     & $\mion$					  	  	  & 1.17					    & 3/2 \mPr \\
\hline 
\end{tabular}
\caption{Free parameters of our stellar model, and our choices of parameters for H-burning stars ($p + p \rightarrow D + e^+ + \nu_e$, as we find in our Universe) and D-burning stars ($p + D \rightarrow \, ^3_2\ro{He}$), which are fueled by an equal mix of hydrogen and deuterium. The polytropic index of stars in our universe is generally believed to increase from $n = 3/2$ for small stars to $n =3$ for high mass stars; the effect on our modelling is small, so we will set $n = 3/2$ for all stars.  The nuclear parameter (\cC) for D-burning stars is calculated from \cite{Krumholz2011}, according to $\cC = 2.1 \ten{17} (X_H / 0.71)^2 (\Deu / 2 \ten{-5})$. The other parameter value choices are explained in Section \ref{Ss:modelparam}.}
\label{tab:table_params}
\end{table*}

The equation of state relates pressure ($P$) to density ($\rho$), $P = K \rho^{\Gamma}$, where $K$ is a constant, and $\Gamma = 1 + 1/n$ is defined in terms of $n$, the polytropic index. We can write the temperature $T = T_c f(\xi)$ and density $\rho = \rho_c f^n(\xi)$ profiles in terms of a dimensionless function ($f$) of radius ($\xi = r/R$), which is a solution of the Lane-Emden equation \cite[][Equation 5]{2008JCAP...08..010A} with boundary conditions $f(0) = 1$, $f'(0) = 0$, and $R^2 = K\Gamma / [(\Gamma - 1)4\pi G \rho_c^{2 - \Gamma}]$. We characterize the length scale of the temperature profile by the dimensionless parameter $\beta$, defined by $f(\beta^{-1}) = e^{-1}$. The outer boundary of the star is given by the solution to $f(\xi_*) = 0$. We can relate these parameters to the total stellar mass $M_*$,
\begin{equation}
M_* = 4 \pi R^3 \rho_c \int_0^{\xi_*} \xi^2 f^n(\xi) \df \xi \equiv 4 \pi R^3 \rho_c \mu_0 ~,
\end{equation}
where we have defined the dimensionless mass parameter $\mu_0$.

Of particular importance is the rate of nuclear energy production, which is given (per unit volume) by,
\begin{equation}
\epsilon(r) = \cC \rho^2 \Theta^2 \exp(-3\Theta) ~,
\end{equation}
where $\cC$ is a constant that summarizes the relevant nuclear physics, $\Theta =  (E_G/4k_B T)^{1/3}$, $E_G = (\pi \alpha Z_1 Z_2)^2 2 \mR c^2$ is the Gamow energy for the reaction, $Z_1$ and $Z_2$ are the electric charges of the reacting nuclei, and $\mR$ their reduced mass.

Defining $\Theta_c = (E_G/4 k_B T_c)^{1/3}$ and $\Theta = \Theta_c f(\xi)^{-1/3}$, the total luminosity is given by integrating the nuclear energy production rate over the whole star: $L_* = \cC 4\pi R^3 \rho_c^2 I(\Theta_c)$, where
\begin{equation}
I(\Theta_c) = \int_0^{\xi_*} f^{2n} \xi^2 \Theta^2 \exp(-3 \Theta) \df \xi ~.
\end{equation}

The central temperature of the star ($T_c$) is determined by the overall thermo- and hydro-dynamical stability of the star; Adams \cite{2008JCAP...08..010A} shows that $\Theta_c$ is a solution to the following equation,
\begin{equation} \label{eq:thc}
I(\Theta_c) \Theta_c^{-8} = \frac{2^{12}\pi^5}{45} \frac{1}{\beta \kappa_0 \cC E_G^3 \hbar^3 c^2} \left(\frac{M_*}{\mu_0}\right)^4 \left(\frac{G \bar{m} }{n+1}\right)^7 ~,
\end{equation}
where $\bar{m}$ is the mean particle mass. We assume that the opacity $\kappa$ varies inversely with the density, so that $\kappa \rho = \kappa_0 \rho_c = $ constant.

Having solved Equation \eqref{eq:thc} to find $\Theta_c$, we can calculate the stellar radius $R_*$, luminosity $L_*$, and surface temperature $T_*$,
\begin{align}
R_* &= \frac{G M_* \bar{m}} {k_B T_c} \frac{\xi_*} {(n+1) \mu_0} \\
L_* &= \frac{16 \pi^4} {15} \frac{1} {\hbar^3 c^2 \beta \kappa_0 \Theta_c} 
\left( \frac{M_*} {\mu_0} \right)^3 \left( \frac{G \bar{m}} {n + 1} \right)^4 \\
T_* &= \left( \frac{L_*} {4 \pi R_*^2 \sigma_{\!_{SB}}} \right)^{1/4}
\end{align}
where $\sigma_{\!_{SB}}$ is the Stefan-Boltzmann constant.

We can give an approximate upper limit for the main-sequence lifetime of a star by calculating the time taken to radiate away the available nuclear energy. We define $\epsilon_{\ro{nuc}}$ to be the ratio of the stars available nuclear energy to the rest mass, which gives,
\begin{equation} \label{eq:tlifetime}
t_* = \frac{\epsilon_{\ro{nuc}} M_* c^2} {L_*} ~.
\end{equation}
Note that this estimate doesn't take into account the effects of stellar evolution, and assumes that the star burns all of the available nuclear fuel.

\subsection{Model Parameters} \label{Ss:modelparam}

We will to use our stellar model to investigate stars that burn via strong (and electromagnetic) interactions, and compare their properties with those that, as in our Universe, include a rate-limiting weak interaction. The third column of Table \ref{tab:table_params} shows the parameters used in \cite{2008JCAP...08..010A} to model stars in our Universe. The polytropic index in our Universe is generally believed to increase from $n = 3/2$ for low mass stars to $n =3$ for high mass stars; the effect on our modelling is small, so we will set $n = 3/2$ for all stars. For simplicity, we assume that our stars are fully ionized throughout.

Rather than change the constants of nature to bind the diproton, we consider a representative of strong-burning stars. Specifically, we model a star in our Universe with initially equal numbers of protons and deuterons. This allows us to use the physical parameters of the D-burning phase in stars, as modelled using Adams's formalism by Krumholz \cite{Krumholz2011}, without having to recalculate nuclear physics from scratch. The deuterium number fraction \Deu is set to unity, from which it follows that the hydrogen mass fraction is $X_H = 1/3$. We calculate the nuclear parameter (\cC) for D-burning stars via $\cC = 2.1 \ten{17} (X_H / 0.71)^2 (\Deu / 2 \ten{-5})$ \cite{Krumholz2011}. We follow \cite{Krumholz2011} by assuming that free electrons are the primary source of opacity: $\kappa_0 = \sigma_{\!_{T}} n_e/\rho = \sigma_{\!_{T}}/\mion$, where $\sigma_{\!_{T}}$ is the electron Thomson cross section, and $\mion = \rho / n_e$ is the mass per electron. We calculate the mass per particle ($\bar{m}$) and mass per electron ($\mion$) via,
\begin{align}
\frac{\bar{m}}{\mPr} &= \frac{\rho}{\mPr n} = \frac{X_H + 2 \Deu X_H + Y_\ro{He}}{2 X_H + 2 \Deu X_H + \frac{3}{4}Y_\ro{He}} \\
\frac{\mion}{\mPr} &= \frac{\rho}{\mPr n_e} = \frac{X_H + 2 \Deu X_H + Y_\ro{He}}{X_H + \Deu X_H + \frac{1}{2}Y_\ro{He}} ~,
\end{align}
where $Y_\ro{He}$ is the helium mass fraction. Unlike Adams, we have not made the simplification that $\bar{m} = \mion$ (though such simplifications make only small changes to our already approximate model).

The proton-proton chain in our Universe begins with the reaction $p + p \rightarrow D + e^+ + \nu_e$, which relies on a slow weak interaction (beta-plus decay). The nuclear burning efficiency is calculated for H-burning stars in our Universe by noting that, in combining 4 protons into $^4$He, 26.73 MeV of energy is released, or 0.0071 of the rest mass energy. The primary deuterium burning reaction is \cite{Krumholz2011},
\begin{align}
^2_1\ro{D} + ^1_1\ro{H} ~ &\rightarrow ~ ^3_2\ro{He} \\
2\, ^3_2\ro{He} ~ &\rightarrow ~ ^4_2\ro{He} + 2\, _1^1\ro{H},
\end{align}
which releases 12.6 MeV per D burned, or 0.0045 of the rest mass energy of the reactants.

\subsection{Making Stable Stars}
Stars of arbitrary size cannot exist in our Universe. If a star is too large, then radiation pressure will dominate over gas pressure in resisting the crush of gravity. The equation of state of the star will approach that of an ultra-relativistic fluid, $P \propto \rho^{4/3}$. Such a system is hydrodynamically precarious: as a Newtonian, self-gravitating body with no rotation, it has neutral stability with respect to radial pulsations. The star can expand or contract with no energy cost \cite{1963MNRAS.125..169H}.

In these circumstances, effects that are usually negligible can determine whether the star is stable, or liable to runaway collapse or mass-loss. The destabilizing effects of general relativity and electron-positron pair formation can be countered by the stabilizing effects of rotation, magnetic fields, and turbulence \cite{Fowler1966,Fuller1986,Baumgarte1999,Goodman2003,Chen2014a}. If stable, these \emph{supermassive} stars are short-lived, burning with luminosity $L \approx 2 \ten{40} (M_* / 100 \Msol)$ erg/s $\approx 10^{7} \Lsol (M_* / 100 \Msol)$ \cite{1963MNRAS.125..169H} before exploding in especially violent supernovae \cite{Chen2014a}. Because $L \propto M_*$, their lifetime of $\sim 10^6$ yrs is roughly independent of their mass. Supermassive stars were originally studied as potential models for quasars \cite{1963MNRAS.125..169H,Fowler1966,Bisnovaty-Kogan1973}. However, radio lobes around quasars show that they can be active for $10^{8}$ years, and thus supermassive stars are too short lived. They are still studied as candidates for Population III stars, and in particular as the seeds for supermassive black holes \cite{Rees1984,Bond1984,Begelman2010,2013ApJ...777...99W,Chen2014a}. Because the lives of these stars are short, marginally stable and violent, we consider the onset of radiation-pressure domination to represent an upper limit to long-lived stars.

Alternatively, if a ball of gas is too small, it will be stabilized by electron degeneracy pressure before gravity can ignite nuclear reactions at its core, resulting in a brown dwarf. Degeneracy pressure depends on the electron number density ($n_e$), which is related to the central mass density by $\mion \equiv \rho_c / n_e$ = constant. With these constraints, Adams \cite{2008JCAP...08..010A} calculates the minimum and maximum masses for stable, long-lived stars.
\begin{align}
M_{*, \ro{min}} &= 6 \sqrt{3 \pi} \left( \frac{4}{5} \right)^{3/4} \left( \frac{m^8_\ro{p}} {\mion^5 \bar{m}^3} \right)^{1/4} \left( \frac{k T_\ro{ign}} {m_e c^2} \right)^{3/4} M_0  \label{eq:Mmin} \\
M_{*, \ro{max}} &= \left( \frac{18 \sqrt{5}} {\pi^{3/2} } \right) 
\left( \frac{1 - f_g} {f_g^4} \right)^{1/2}  \left( \frac{m_\ro{p}}{\bar{m}} \right)^2 \, M_0 \label{eq:Mmax}
\end{align}
where $T_\ro{ign}$ is the \emph{ignition temperature}, that is, the minimum central temperature that a star needs to sustain nuclear reactions; $f_g$ is the fraction of the star's central pressure provided by the gas. We assume that $f_g = 1/2$ marks the onset of radiation-pressure dominated, short-lived stars.

Using the relationship between stellar mass and central stellar temperature in Equation \eqref{eq:thc}, we can use the expression for $M_{*, \ro{min}}$ \eqref{eq:Mmin} to calculate $T_\ro{ign}$ for stars in a given universe, that is, for a given set of fundamental constants. The corresponding $\Theta_c$ is the solution to the following equation,
\begin{equation} \label{eq:Tminnuc}
J(\Theta_c) \equiv \Theta_c I(\Theta_c) = \left( \frac{2^{16} \pi^7 3^4} {5^{4} } \right) 
\left( \frac{\hbar^3} {c^2} \right) \left( \frac{1} {\beta \mu_0^4} \right) 
\left( \frac{\bar{m}^4} {\mion^5 m_e^3} \right) \left( \frac{G} {\kappa_0 \cC (n + 1)^7} \right)
\end{equation}
where we have defined the function $J(\Theta_c)$. Using our preferred parameters for H-burning stars from Table \ref{tab:table_params}, we find that $M_{*, \ro{min}} = 0.18 \Msol$ and $M_{*, \ro{max}} = 102 \Msol$, which is in reasonable agreement with the range of stellar masses observed in our Universe.

\subsection{Stable Stars in Parameter Space}

The window of possible stellar masses -- between $0.18 \Msol$ and $102 \Msol$ in our Universe --- can vanish in other parts of parameter space. With other fundamental constants, a ball of gas which is large enough to ignite is too large to live a long, stable life. We can determine where the stellar window vanishes by equating the minimum and maximum stellar masses in Equations \eqref{eq:Mmin} and \eqref{eq:Mmax}. This gives the maximum ignition temperature, $T_\ro{ig,max}$.
\begin{equation} \label{eq:Tmax}
\frac{kT_\ro{ig,max}} {m_\ro{e} c^2} = 
\left( \frac{3^2 5^5}{2^6 \pi^8} \right)^{1/3}
\left( \frac{1 - f_g}{f_g^4} \right)^{2/3}
\left( \frac{\mion}{\bar{m}} \right)^{5/3} ~.
\end{equation}
If the ignition temperature in a given universe ($T_\ro{ign}$) is greater than $T_\ro{ig,max}$, then stable stars are not possible --- anything large enough to ignite nuclear reactions is unstable. The right hand side is $\approx 4$ for our models, corresponding to $T_\ro{ig,max} = 2.5 \ten{10}$ K. This is well above typical hydrogen burning temperatures in our Universe ($10^7$ K). From $T_\ro{ig,max}$, we can calculate $\Theta_\ro{ig}$.

\begin{figure*} \centering
	\begin{minipage}{0.45\textwidth}
		\includegraphics[width=\textwidth]{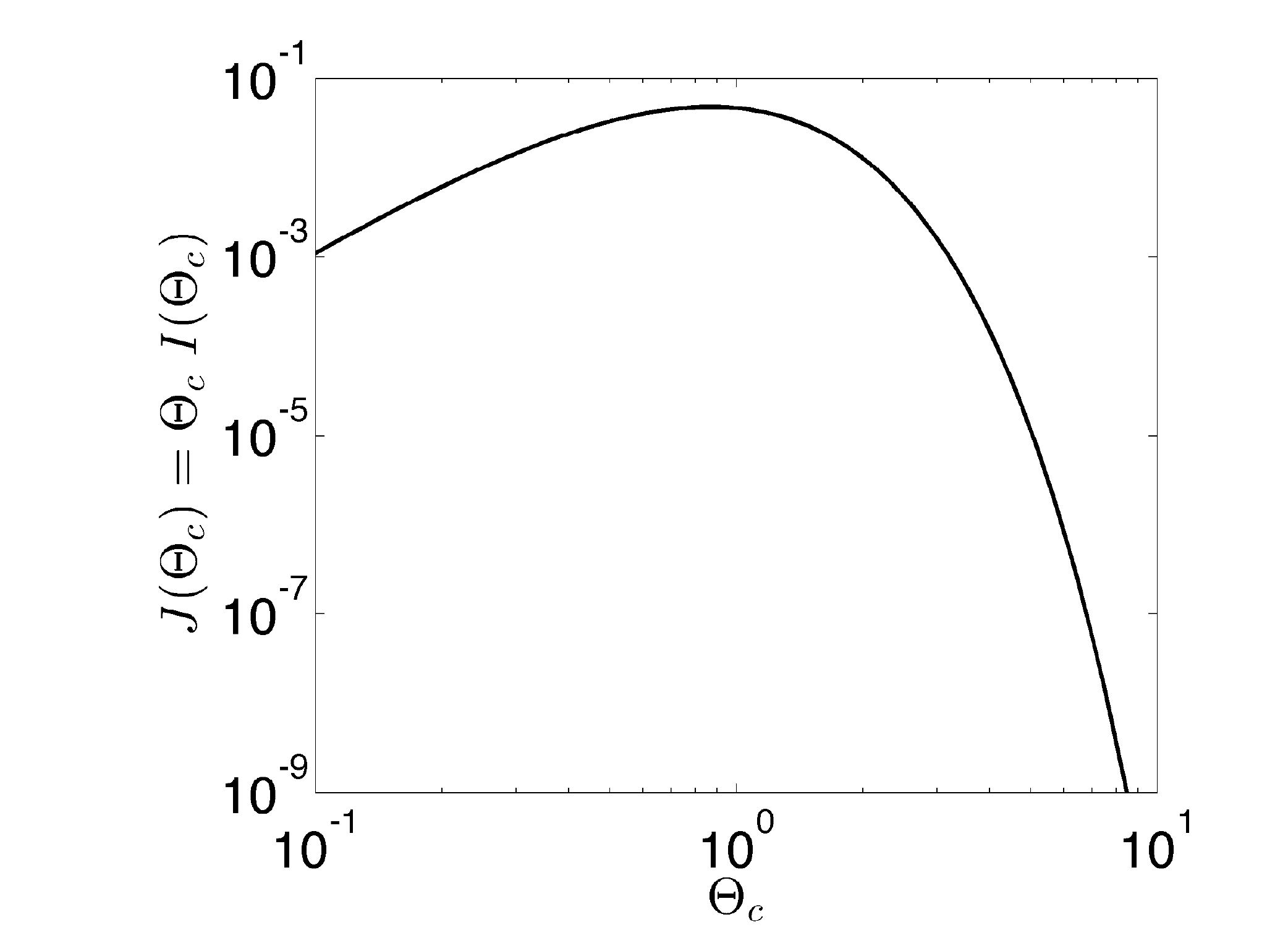}
	\end{minipage}
	\begin{minipage}{0.45\textwidth}
		\caption{The left hand side of Equation \eqref{eq:Tminnuc} $J(\Theta_c) \equiv \Theta_c I(\Theta_c)$, plotted as a function of $\Theta_c = (E_G/4 k_B T_c)^{1/3}$. $J(\Theta_c)$ has a maximum of $J_\ro{max} = 0.0478$ at $\Theta_\ro{max} = 0.869$.}
		\label{fig:ITheta}
	\end{minipage}
\end{figure*}

Recall that the minimum central temperature for stars is given by the solution to Equation \eqref{eq:Tminnuc}. The left hand side of the equation [$J(\Theta_c) \equiv \Theta_c I(\Theta_c)$] has a maximum of $J_\ro{max} = 0.0478$ at $\Theta_\ro{max} = 0.869$ (Figure \ref{fig:ITheta}). For certain values of the fundamental constants, Equation \eqref{eq:Tminnuc} has two solutions. The physical solution is the one with the lower temperature, which corresponds to a larger $\Theta_c$.

There are thus two ways that a universe can fail to make stable stars. If $\Theta_\ro{ig} < \Theta_\ro{max}$, then the unstable ignition temperatures of Equation \eqref{eq:Tmax} are unphysical anyway, so we need only worry that Equation \eqref{eq:Tminnuc} has a solution at all (i.e. that the right hand side is less than $J_\ro{max}$). Alternatively, if $\Theta_\ro{ig} > \Theta_\ro{max}$, then stable stars require that the solution to Equation \eqref{eq:Tminnuc} has $\Theta_c > \Theta_\ro{ig}$, so that $T_c < T_\ro{ig,max}$. We can combine these cases by calculating the maximum value of the gravitational constant $G$ consistent with stable stars, rearranging Equation \eqref{eq:Tminnuc},
\begin{equation} \label{eq:Gmax}
G_\ro{max} = J( \ro{max}\{\Theta_\ro{ig},\Theta_\ro{max} \})
\left( \frac{5^4} {2^{16} \pi^7 3^4} \right) 
\left( \frac{c^2} {\hbar^3} \right)
\left( \frac{\mion^5 m_e^3} {\bar{m}^4} \right) \beta \mu_0^4 \kappa_0 \cC (n + 1)^7 ~.
\end{equation}
We follow Adams \cite{2008JCAP...08..010A} by also varying the fine-structure constant, as it affects the stellar opacity $\kappa_0 \propto \sigma_{\!_{T}} \propto \alpha^2$ and the Gamow energy $E_G \propto \alpha^2$; note that changing $E_G$ affects $\Theta_\ro{max}$\footnote{Davies \cite{Davies1972} shows that if $\alpha_s/\alpha_{s,0} \gtrsim 1.003 + 0.031 \alpha/\alpha_0$, where $\alpha_s$ is the strong coupling constant and a subscript zero refers to the value of the constant in our Universe, then the diproton is unbound. Since $\alpha$ is positive, there is no value of the fine-structure constant that will bind the diproton, given the strength of the strong force in our Universe. This can be seen from the fact that the dineutron is unbound, so that adding extra internal Coulomb repulsion makes the diproton more unstable.}.

\section{Deuterium Burning Stars} \label{S:Dburn}

\subsection{Stellar Models}
\begin{figure*} \centering
	\begin{minipage}{0.45\textwidth}
		\includegraphics[width=\textwidth]{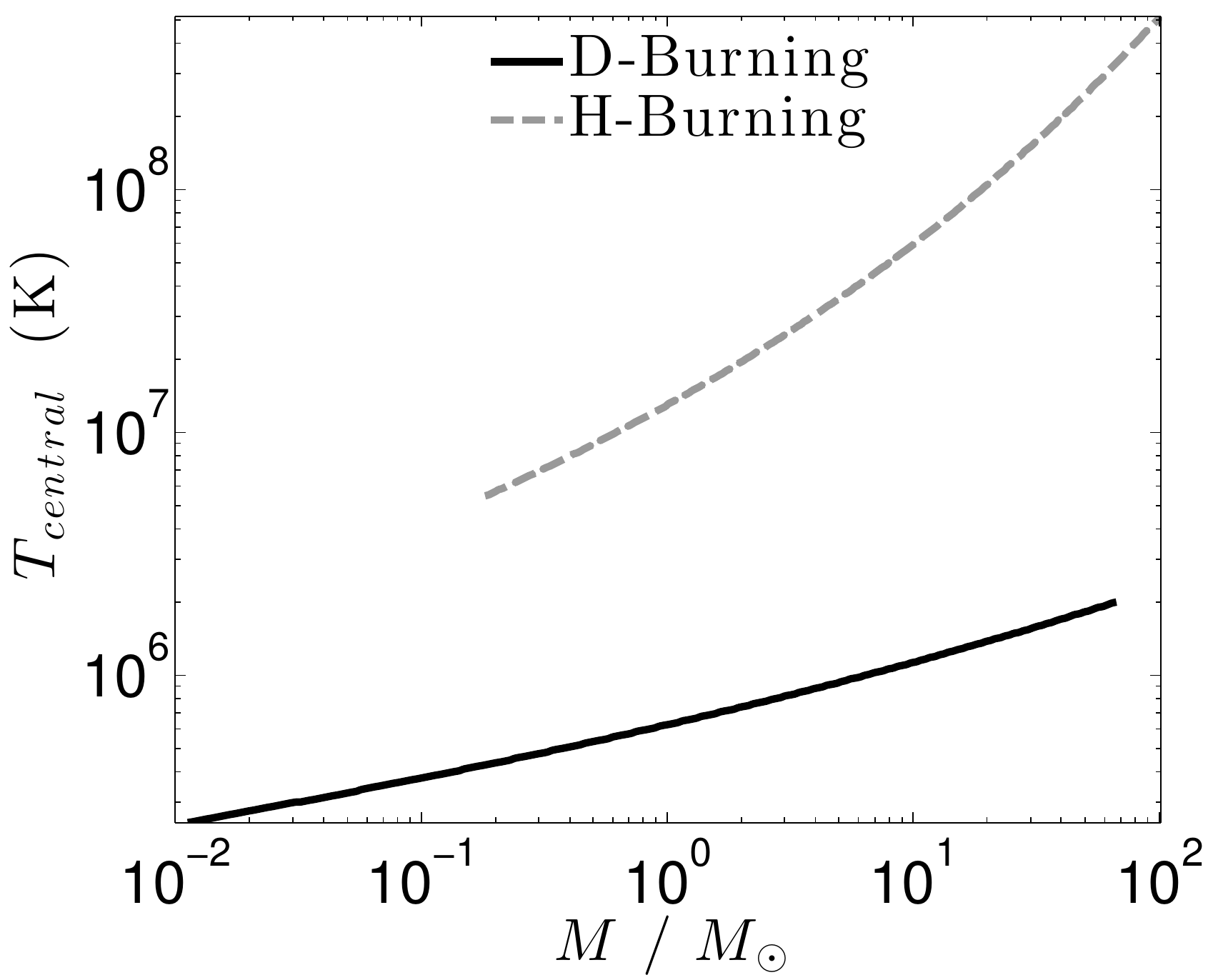}
	\end{minipage}
	\begin{minipage}{0.45\textwidth}
		\includegraphics[width=\textwidth]{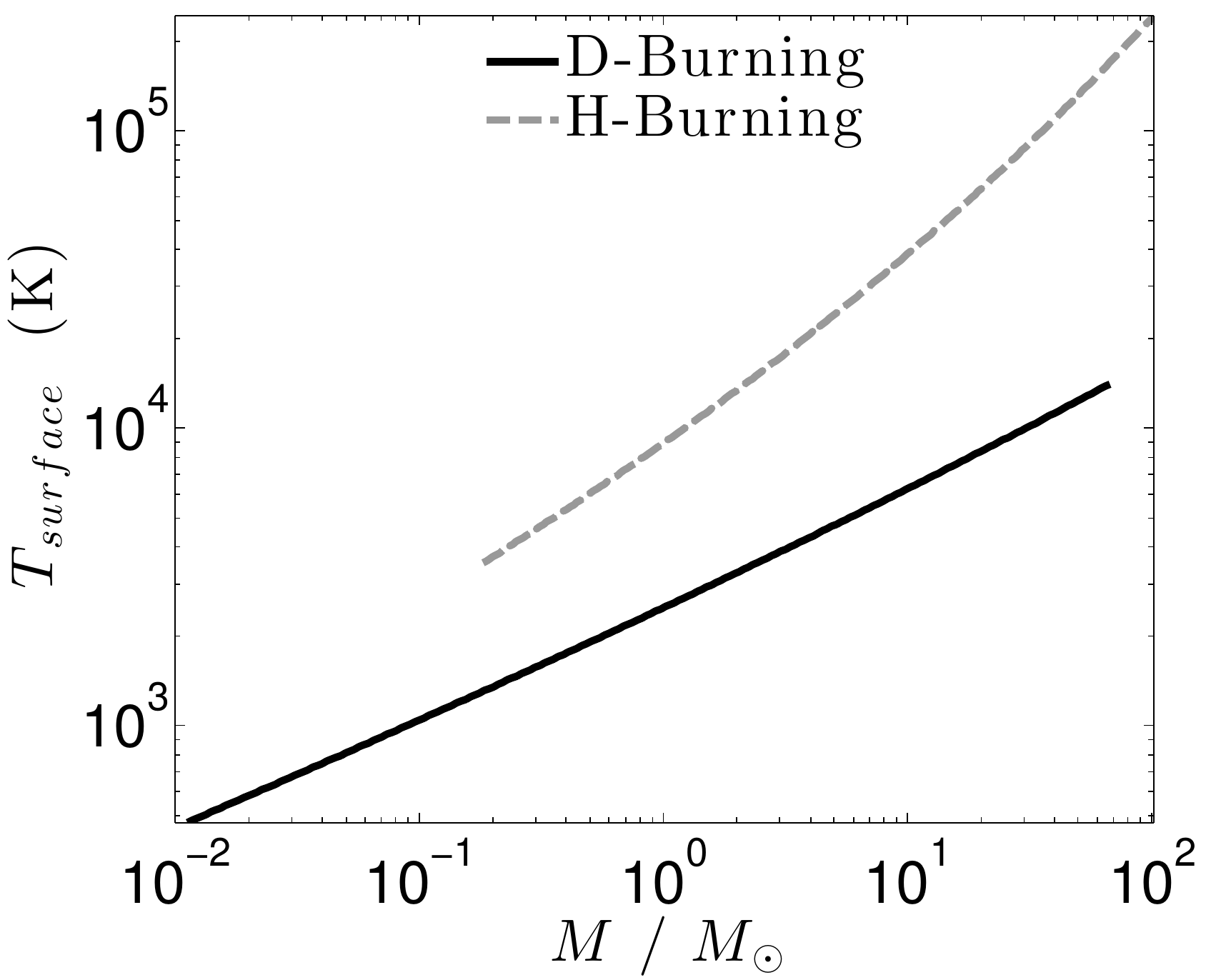}
	\end{minipage}
	\begin{minipage}{0.45\textwidth}
		\includegraphics[width=\textwidth]{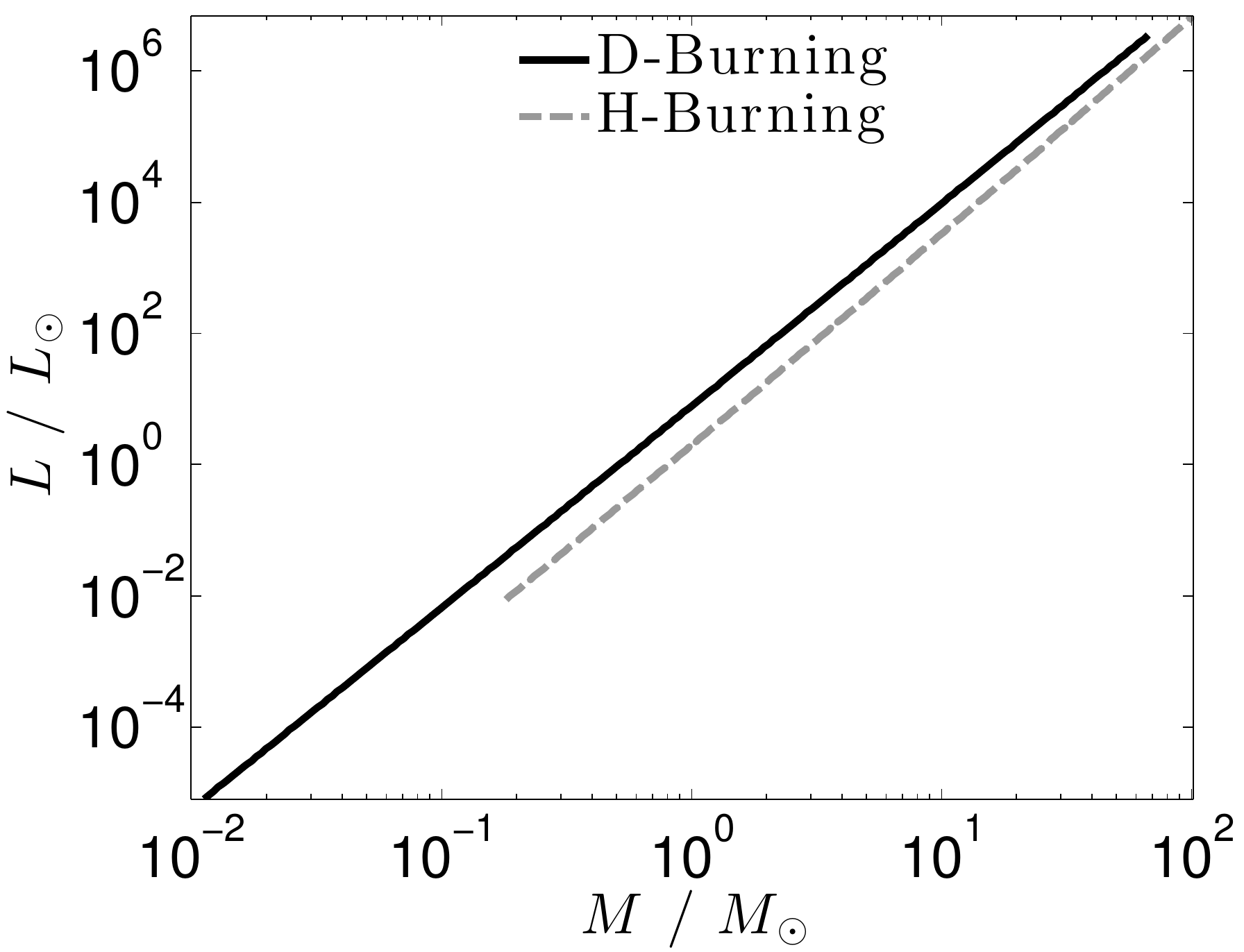}
	\end{minipage}
	\begin{minipage}{0.45\textwidth}
		\includegraphics[width=\textwidth]{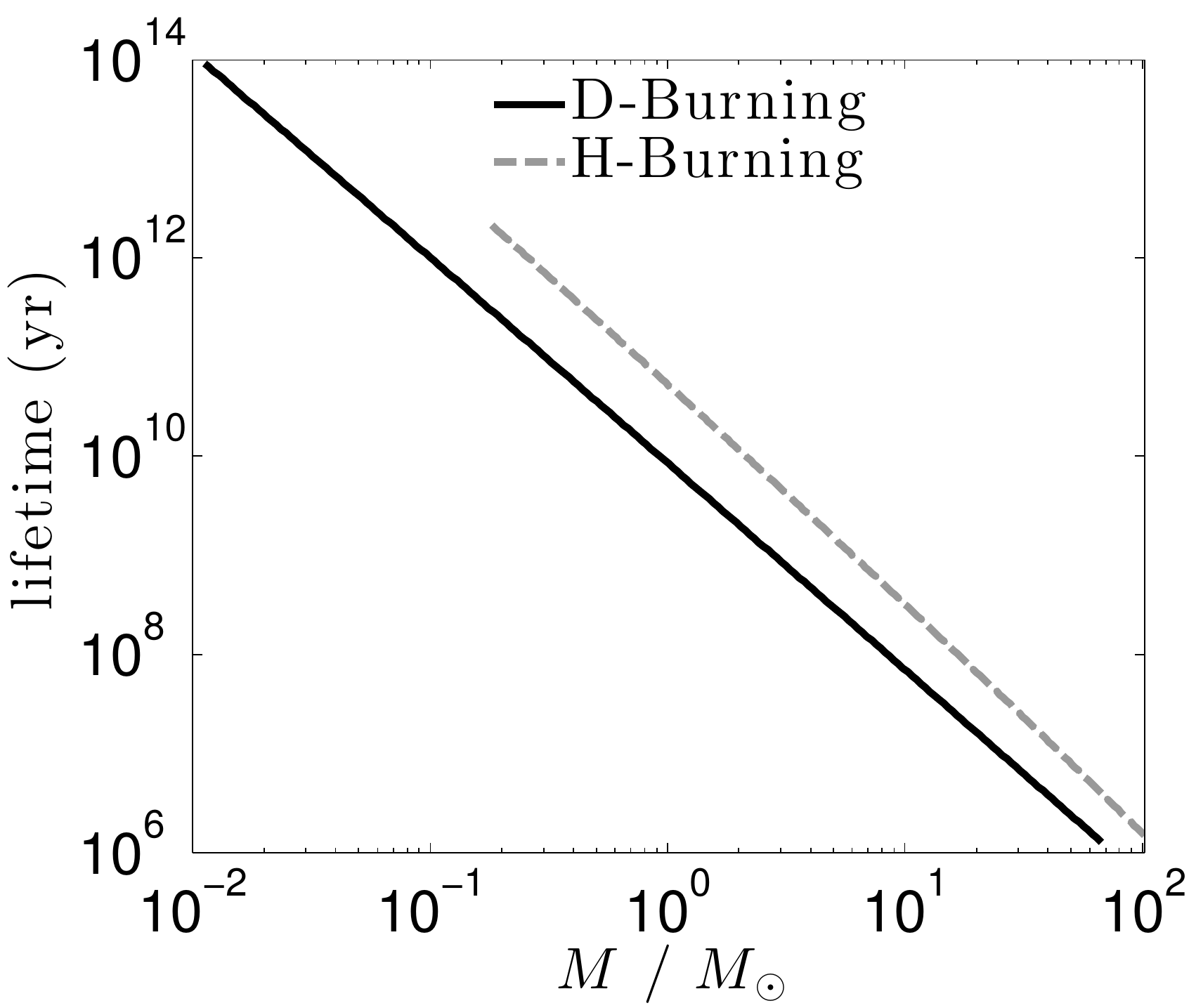}
	\end{minipage}
\caption{The properties of D-burning  ($\Deu = 1$, black line) and H-burning (grey line) stars. The panels show the relationship between stellar mass and core temperature (top left), surface temperature (top right), luminosity (bottom left), and lifetime (top right). }
\label{fig:Dmodels}
\end{figure*}

We now use our stellar model to calculate the properties of D-burning stars ($\Deu = 1$). Figure \ref{fig:Dmodels} shows the relationships between the fundamental stellar properties of mass, core temperature, surface temperature, luminosity, and lifetime. 

The range of permitted masses for D-burning stars is $0.01 \Msol$ to $67 \Msol$, which overlaps significantly with H-burning stars. The reason that D-burning stars can be $\sim 10$ times less massive that H-burning stars is their lower ignition temperatures, as shown in the top left panel. The central temperature of D-burning stars is 1-2 orders of magnitude lower than that of H-burning stars, while their central density (not shown) is 4-6 orders of magnitude lower. The net result is that the central nuclear reaction rate of D-burning stars is actually several orders of magnitude lower than for H-burning stars, in spite of \cC being 17-18 orders of magnitude larger.

The top right plot of Figure \ref{fig:Dmodels} shows stellar surface temperature ($T_\ro{surface}$). The difference in $T_\ro{surface}$ between the of D- and H-burning stars is less than an order of magnitude, with a large overlap in the ranges of $T_\ro{surface}$. The surface temperature of a star determines the typical energy of its emitted photons. H-burning stars emit photons with similar energy to molecular bonds, which is important to biological processes such as photosynthesis. Many D-burning stars share this property.

Note that $T_\ro{surface}$ for H-burning stars is somewhat overestimated; for example, the temperature of a star with the mass of the Sun is $\sim 50 \%$ larger than $T_\odot = 5,778$ K. This is partly due to assuming $n = 3/2$ for all stars. Setting $n=3$ for the Sun gives result $T_\ro{surface} \approx 5200$ K, which is within 10\% of the correct value. We could follow Adams \cite{2008JCAP...08..010A} and vary $n$ with stellar mass to improve our model's predictions of H-burning stars in this Universe, but it is not clear what mass-$n$ relationship we should assume for D-burning stars. We use $n = 3/2$ as this is more accurate for low-mass stars.

The bottom left panel of Figure \ref{fig:Dmodels} shows stellar luminosity ($L_*$), and demonstrates the remarkable similarity between the two types of star. Over eight orders of magnitude in luminosity, only about a factor of three separates the luminosities at a given mass. The bottom right panel of Figure \ref{fig:Dmodels} shows approximate stellar lifetime ($t_*$) vs mass. While at a given stellar mass, H-burning stars live 5-10 times longer, very long lived ($t_* > 10$ Gyr) D-burning stars are stable.

In summary, D-burning stars with familiar surface temperatures, luminosities, and lifetimes are available. At worst, a D-burning star with a similar lifetime to, say, the Sun would have a smaller mass and thus a lower surface temperature, perhaps $\sim 1,000$ K. Nearby planets would be bathed in near infrared radiation, which is not a definitively life-prohibiting outcome. 

\subsection{Strong-Burning Stars in Parameter Space}

\begin{figure*} \centering
	\begin{minipage}{0.45\textwidth}
		\includegraphics[width=\textwidth]{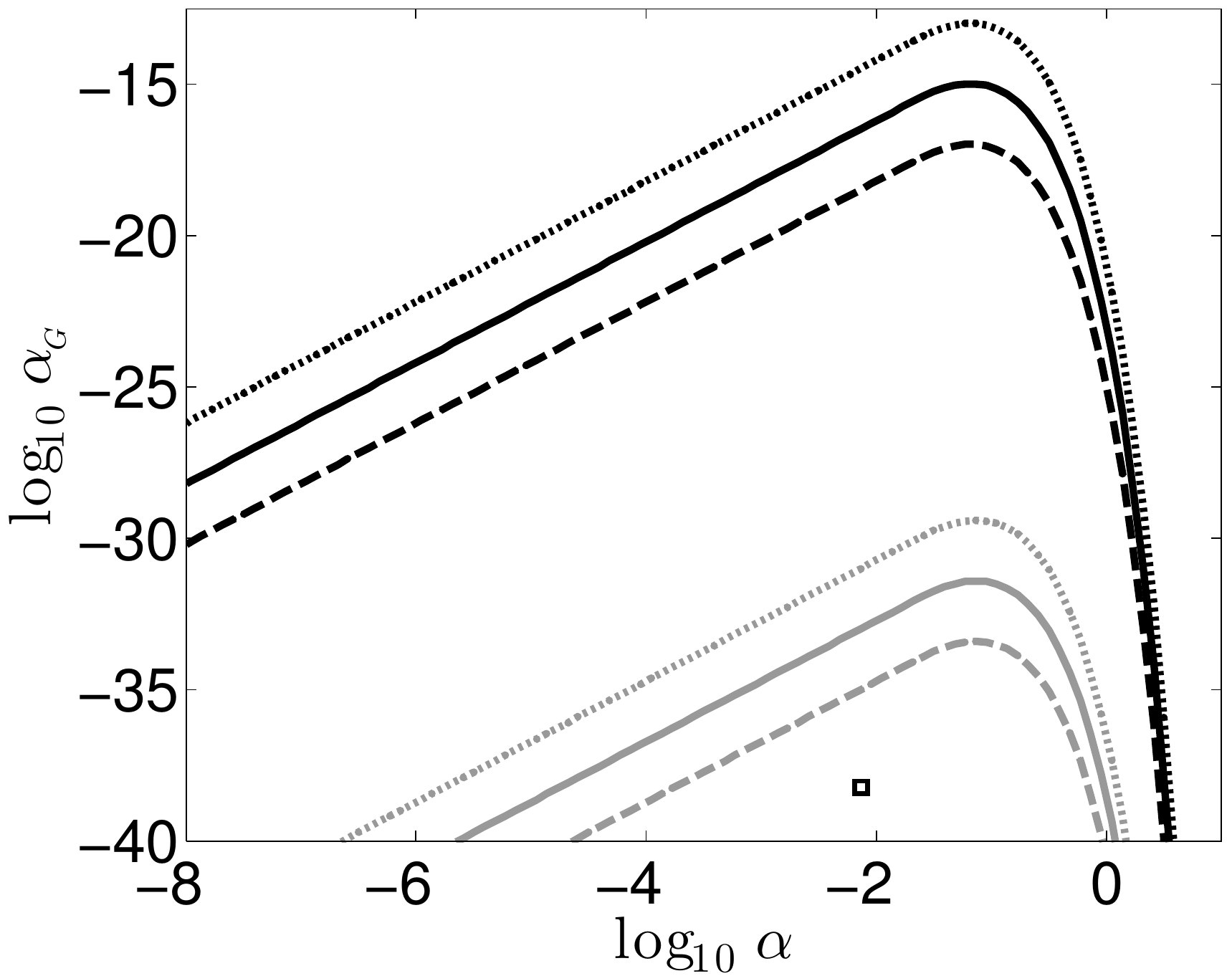}
	\end{minipage}
	\begin{minipage}{0.50\textwidth}
		\caption{Limits on stable stars in a 2-dimensional slice through parameter space, $\alG = \mPr^2 / m^2_\ro{Planck}$ vs. $\alpha$ for H-burning stars (black lines) and D-burning stars (grey lines). The black square shows our Universe's constants. The dotted (dashed) lines show the effect of increasing (decreasing) the nuclear burning parameter \cC by a factor of 100. Stars that burn via the strong force, such as D-burning stars, are stable in a much larger region of parameter space.}
		\label{fig:DH_lim}
	\end{minipage}
\end{figure*}

Having seen that H-burning and D-burning stars can have similar properties, we turn to the place of stable stars in parameter space. The effect of increasing the nuclear reaction parameter by $\sim 17-18$ orders of magnitude can be seen in Equation \ref{eq:Gmax}, where $G_\ro{max}$ increases in proportion with \cC. Stable stars that burn via the strong and electromagnetic force thus occupy a substantially larger portion of parameter space than stars that, as in our Universe, are slowed by the weak interaction. The relevant limits are shown in Figure \ref{fig:DH_lim}, plotted against the dimensionless parameters $\alpha$ and $\alG$ for D-burning stars (black lines) and H-burning stars (grey lines)\footnote{When the electromagnetic coupling constant ($\alpha$) becomes large ($\gtrsim 1$), QED becomes a strongly-coupled theory. It cannot be approached perturbatively, which makes calculation significantly more difficult. The familiar weak-coupling or \emph{Coulomb phase} of QED, with its $1/r$ potential and massless photon, transitions to a \emph{confinement phase}. The photon becomes massive, and the influence of magnetic monopoles leads to a linear potential ($\propto r$). Just as quarks are confined by QCD in our Universe, so charges are confined in strongly-coupled QED \cite{Smit2002,Laperashvili2001,Bozkaya2004}. Thus, the region of Figure \ref{fig:DH_lim} and following figures where $\alpha \gtrsim 1$ probably represents an unjustified extrapolation of electromagnetism into a qualitatively different, and difficult to calculate, regime.}. Changing the value of $G$ is equivalent to scaling the values of all the relevant particle masses. For this reason, we express changes in $G$ in terms of the dimensionless parameter $\alG = (\mPr/m_\ro{Planck})^2$. The black square shows our Universe's constants.

The dotted (dashed) lines in Figure \ref{fig:DH_lim} show the effect of increasing (decreasing) the nuclear burning parameter \cC by a factor of 100. Stars that burn via the strong force, such as D-burning stars, are stable in a much larger region of parameter space. The maximum value of $\alG$ which, regardless of $\alpha$, can support stable H-burning stars is $\sim 10^7$ times greater than its value in our universe, while D-burning stars could be stable if $\alG$ were increased by $\sim 10^{23}$.

\subsection{Stellar Ages and Surface Temperature}
There are, however, other reasons to not regard the entirety of the parameter region opened up by stable strong force stars as life permitting. We can also consider the lifetime of the star, which should be long enough that nearby planets are provided with energy on timescales over which biological evolution can progress. We can also consider the surface temperature of stars. Life on Earth relies on solar photons that are energetic enough to excite electrons and power chemical reactions, but without being so energetic that they destroy living tissue and sterilize the planetary surface. \cite{1983RSPTA.310..323P} shows that the coincidence between the energy of stellar photons and the energy of molecular bonds relies on the following numerical coincidence between fundamental constants: $\alG^{1/8} \approx \alpha^{3/2} (m_\ro{e}/m_\ro{p})^{1/2}$.

\begin{figure*}[t] \centering
	\begin{minipage}{0.49\textwidth}
		\includegraphics[width=\textwidth]{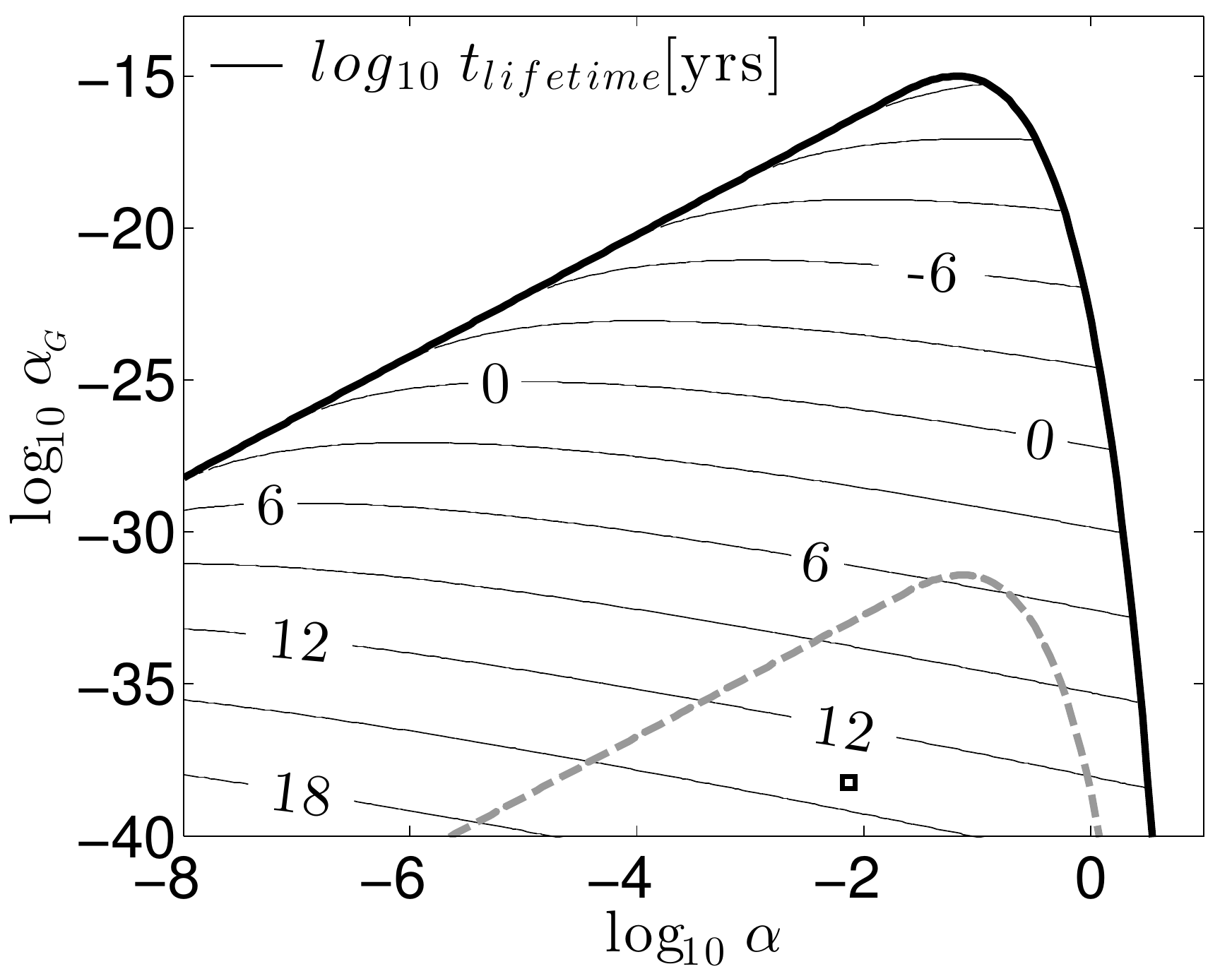}
	\end{minipage}
	\begin{minipage}{0.49\textwidth}
		\includegraphics[width=\textwidth]{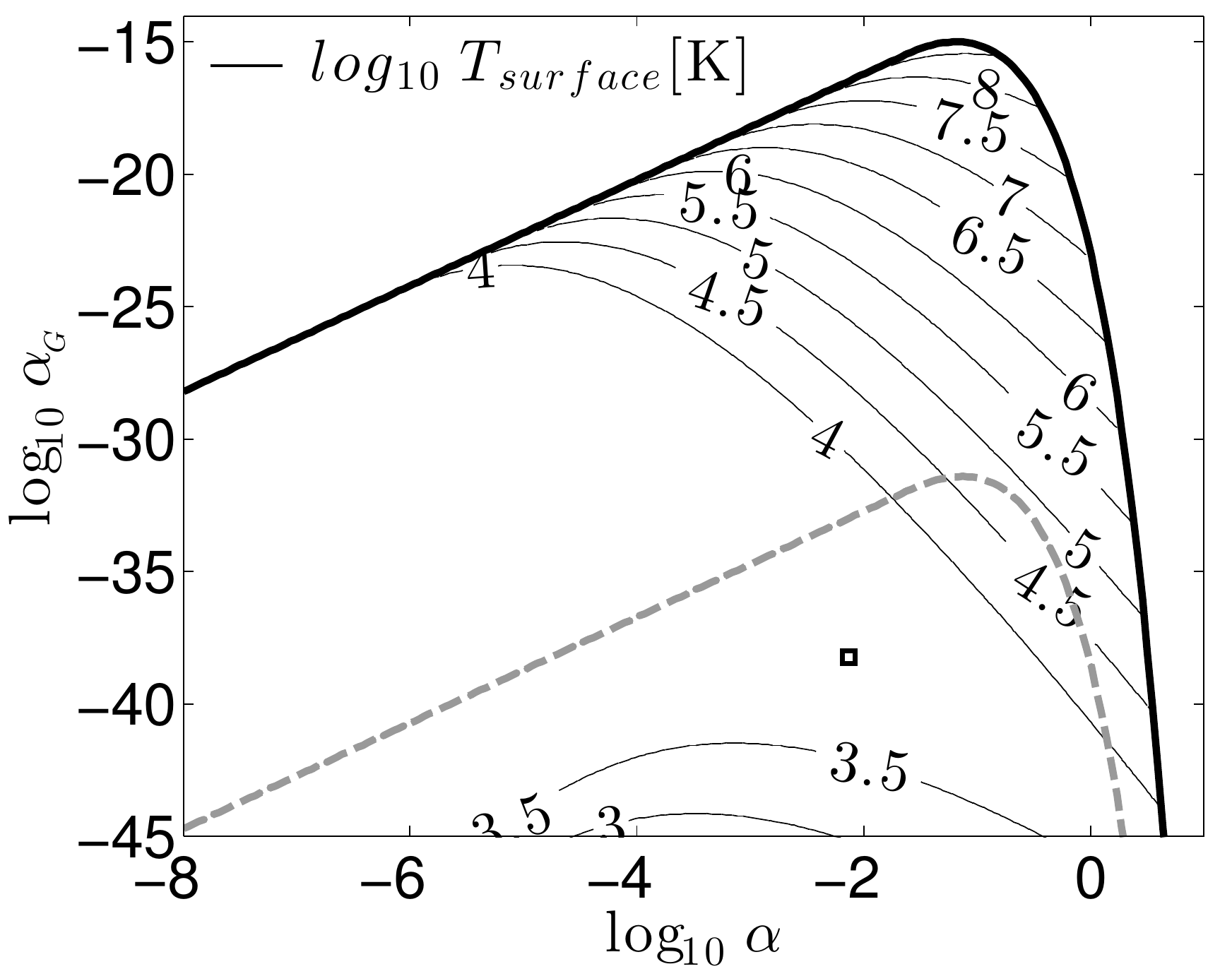}
	\end{minipage}
		\caption{Similar to Figure \ref{fig:DH_lim}, but overlaid with contours of stellar age (left) and surface temperature (right) on the region of parameter space over which D-burning stars are stable. The black solid (grey dashed) lines show the stability limits for D-burning (H-burning) stars. Specifically, the contours on the left panel are labeled with the age of the \emph{longest lived} (and hence, least massive) star possible in a universe with a given set of constants. The contours on the right panel are labeled with the surface temperature of the star whose temperature is closest to that of the Sun. The contours above our Universe (black square) show where the coolest stable star is hotter than the Sun, while the contours below show where the hottest stars are cooler than the Sun.}
		\label{fig:DH_agesurf}
\end{figure*}

Figure \ref{fig:DH_agesurf} shows contours of stellar age and surface temperature on the region of parameter space over which D-burning stars are stable. The black solid (grey dashed) lines show the stability limits for D-burning (H-burning) stars, as shown in Figure \ref{fig:DH_lim}. The black square shows our Universe's constants.

The contours on the left panel of Figure \ref{fig:DH_agesurf} are labeled with the age of the \emph{longest lived} (and hence, least massive) star possible in a universe with a given set of constants. Much of the region of parameter space that has been opened up contains very short-lived stars. For example, requiring that \emph{some} star in a given universe can live longer than a million years limits the gravitational coupling constant to be $\alG \lesssim 10^{30}$, regardless of the value of the fine-structure constant. Remember, also, that these ages are probably overestimated by Equation \ref{eq:tlifetime}, as it assumes that the star will consume all available fuel. The Sun, but contrast, is only expected to consume about 10\% of its available hydrogen over its life.

The contours on the right panel of Figure \ref{fig:DH_agesurf} are labeled with the surface temperature of the star whose temperature is closest to that of the Sun. The contours above our Universe (black square) show where the coolest stable star is hotter than the Sun, while the contours below show where the hottest stars are cooler than the Sun. Holding $\alpha$ constant, stars with temperatures above $\sim 10^5$ K emit photons with typical energies in excess of atomic ionization energies, which are very damaging to biological organisms. Photons of this energy are recommended by the U.S. Environmental Protection Agency for use in germicidal UV lamps \cite{EPA2006}. Comparing with the left panel, many of these energetic stars burn for less than a million years.

\section{Discussion} \label{S:Discuss}
On the basis of our models of stars that burn by the strong and electromagnetic interactions, as opposed to those whose nuclear reactions include a rate-limiting weak interaction, we conclude that the binding of the diproton is not a disaster. Rather than shining $10^{18}$ times brighter than the Sun, strong-burning stars (such as D-burning stars) can have remarkably familiar masses, luminosities, surface temperatures and lifetimes. Given also that the isotopes of hydrogen are not necessarily burned to helium in the first few minutes of the universe \cite{2009JApA...30..119B,2009PhRvD..80d3507M}, the binding of the diproton does not represent a decisive anthropic boundary in parameter space.

\bigskip\noindent
\emph{Anthropic Constraints on the Strength of Gravity.} In a universe in which stars burn via electromagnetic and strong interactions, the most important constraint on life-permitting stars comes from their lifetime, which translates into a constraint on the strength of gravity. In universes in which gravity is stronger by $\sim$9--10 orders of magnitude, all stars burn out in mere millions of years, regardless of the value of the fine-structure constant.

Should we think of this stable-star-permitting region of parameter space as small or large? The smallness of $\alG = 5.9 \ten{-39}$ in our Universe reflects the smallness of the proton mass with respect to the Planck mass. This in turn reflects the smallness of the contributions to the proton mass. Why is the Higgs vacuum expectation value ($v = 246$ GeV), which gives the quarks their mass, so much smaller than the Planck scale? This problem ``is so notorious that it's acquired a special name --- the Hierarchy Problem --- and spawned a vast, inconclusive literature'', says Wilczek \cite{Wilczek2006a}. (Moreover, why are the light quark and electron masses so much smaller than $v$? Or, equivalently, why are their Yukawa parameters $\sim 10^{-6} \ll 1$?) Also, why is the QCD mass scale ($\Lambda_\ro{QCD} \approx 200$ MeV) so much smaller than the Planck scale? The logarithmic running of the QCD coupling constant means that grand unified theories may be able to naturally explain the smallness of $\Lambda_\ro{QCD}$ \cite{2003AnHP....4..211W}.

In light of this, the constraint of $\alG \lesssim 10^{-30}$ from Figure \ref{fig:DH_agesurf} implies a strong anthropic reason for the fundamental mass scales of the standard model to be at least 15 orders of magnitude smaller than the Planck scale. Note that the relevant comparison is not of the permitted change in the constant with the value in our Universe, since our value of $\alG$ is known to be unnaturally small. Rather, we compare the life-permitting range of a parameter with the range over which its effects can be reliably predicted. For a mass parameter, and in the absence of a quantum theory of gravity, this is from zero to (at most) the Planck mass.

\bigskip\noindent
\emph{Anthropic Constraints on the Light Quark Masses.} As noted in the introduction, avoiding the ``diproton disaster" places very tight anthropic constraints on the masses of the lightest quarks. Given that the binding of the proton is not as disastrous as previously believed, what now are the tightest anthropic constraints on the sum of the light quark masses?

Recent calculations of the properties of the triple alpha process in stars \cite{Epelbaum2013} have demonstrated that astrophysically significant changes to the energy of the Hoyle resonance ($\sim$100 keV) require a $2-3$\%  shift in the sum of the masses of the light quarks ($m_\ro{up} + m_\ro{down}$). A shift of 277 keV radically changes the results of the later stages of stellar evolution, resulting in very little carbon being produced \cite{1989Natur.340..281L}. The changes in the light quark masses that bind the diproton (decrease by $\sim 25$ \%) or strongly unbind the deuteron (increase by $\sim 40$ \%) are likely to result in an effectively carbon- and/or oxygen-free Universe. While not as spectacularly life-prohibiting as a diproton-burning Sun that exhausts its entire fuel supply in a few seconds, the Hoyle resonance limits are anthropically significant and place constraints on the sum of the quark masses that are similar to the ``diproton disaster" limits.

\acknowledgments
Supported by a grant from the John Templeton Foundation. This publication was made possible through the support of a grant from the John Templeton Foundation. The opinions expressed in this publication are those of the author and do not necessarily reflect the views of the John Templeton Foundation.


\end{document}